\documentclass[12pt]{iopart}
\usepackage{graphicx}
\usepackage{dcolumn}

\begin{document}


\title{Optical absorption of semiconductor 2D Majorana nanowires} 



\author{Daniel Ruiz$^1$, Javier Osca$^1$ and L. Serra$^{1,2}$}
\address{Institut de F\'{\i}sica Interdisciplin\`aria i de Sistemes Complexos
IFISC (CSIC-UIB), E-07122 Palma de Mallorca, Spain}
\address{Departament de F\'{\i}sica,
Universitat de les Illes Balears, E-07122 Palma de Mallorca, Spain}
\ead{druiz@ifisc.uib-csic.es, javier@ifisc.uib-csic.es, llorens.serra@uib.es}

\date{February 25, 2014}

\begin{abstract}

We calculate the cross section for optical absorption of planar 2D Majorana nanowires. Light is described 
in the dipole approximation. We discuss the signatures on the cross section of a near-zero-energy
mode. A low energy peak for transverse polarization, absent in longitudinal one, 
reveals the presence of the Majorana-like state. This peak is relatively robust against
thermal smearing of the level occupations. We consider the influence of optical masks hiding parts
of the nanowire from the light.

\end{abstract}


\maketitle 
\section{Introduction}
\label{intro}

Majorana suggested in 1937 the idea of fermionic particles that are 
their own antiparticles \cite{Majorana}.
In spite of many years of research,
especially on neutrino-less double beta decay, the 
existence of elementary particles of Majorana character is still unclear.
Recently, however, in condensed matter physics the concept of Majorana states 
has attracted much attention, partly due to the connections with particle physics
and partly due to the possible applications in quantum computation \cite{Nayak,Alicea,Beenakker,Franz,StanescuREV}.

In condensed matter physics a Majorana mode is a collective state that emerges due to many-body interactions and whose properties resemble those of Majorana elementary particles.
The existence of such quasiparticles has been predicted theoretically in 
several topological condensed matter systems \cite{Qi,Hasan}.
In a hybrid semiconductor-supercoductor nanowire a Majorana mode is formed when the lowest energy levels collapse to zero and the fermionic states fuse in a unique state characterized by a wave function localized on the nanowire  ends. Three mechanisms are needed to form these zero modes
in hybrid nanowires: superconductivity, Rashba spin-orbit coupling and Zeeman magnetic 
field \cite{Fu,Lutchyn,Oreg,Klino2,Flensberg,Akhmerov}.

One of the reasons of the interest in Majorana modes is the fact that they are 
non-abelian anyons, i.e., their state changes in a non-trivial way when two such quasiparticles are interchanged. Moreover, due to their localized character Majorana modes are topologically protected against decoherence, e.g., as induced by sources of noise. It is believed that because of these features Majorana states are good candidates to be used in future quantum computers \cite{Nayak}.

Recent measurements of the electrical conductance of hybrid nanowires 
have provided good 
evidences on the existence of Majorana states in these systems \cite{Mourik,Deng,Rokhinson,Das,Finck}. However, these evidences are not enough to unambiguously confirm the existence of the Majorana states
(discarding for instance similar physics originating from disorder, smooth confinement, Kondo effect or Andreev states) \cite{Pientka,Kells,Lee1,Lee2}.
As more evidences on Majorana modes are presently looked for in the community,
it has been suggested to consider the coupling with the electromagnetic field in
photonic cavities \cite{Trif,Cottet}. 
In a related direction, we explore in this 
work the simpler 
optical absorption of 2D hybrid nanowires, focussing on the signatures of the existence of
a near-zero energy mode in the system. We use the dipole approximation to describe the optical 
field and consider the linear response formalism to a weak perturbation \cite{Merzbacher}.
Similar formalisms for the absorption by quasiparticles in superconductors 
can be found in Refs.\ \cite{Janko,Plehn,Rosen}. 

We find that 
for field polarization parallel
to the nanowire (see Fig.\ \ref{F2.1}) the Majorana state leaves no clear signature on the 
cross section. By contrast,
the existence of the Majorana is indicated by a low energy 
peak emerging for transverse ($\hat{y}$) polarization. This low energy feature is relatively robust 
when the temperature is increased and many levels become thermally activated due to the change
in occupations. 
We also considered the influence of optical masks, hiding parts of the nanowire to the optical excitation,
as a probe of the localization character of the Majorana modes.
Based on these results, we believe that in the characterization 
of Majorana states in nanowires optical absorption experiments 
would be relevant.

\section{Physical model}
\label{sec:1}

\begin{figure}[t]
\begin{center}
	\includegraphics[scale=0.6]{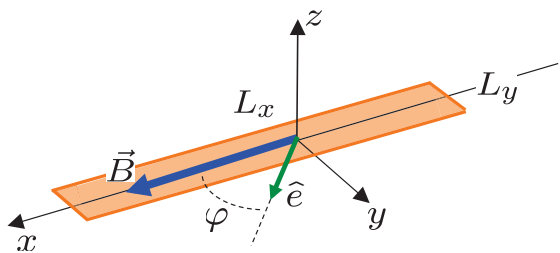} 
	\caption{Sketch of the planar nanowire in magnetic field. The polarization vector of a dipole field for an azimuthal angle $\varphi$ in the $xy$ plane is indicated.}
	\label{F2.1}
\end{center}
\end{figure}

We model a semiconductor nanowire with spin-orbit Rashba coupling, in the presence of a magnetic field, while a nearby superconductor induces the superconductivity effect due to proximity.
The system is described with a Bogoliubov-deGennes Hamiltonian
\begin{eqnarray}
\label{eq.2.1}
\mathcal{H}_{BdG}
&=&
\left[\frac{p^2_x+p^2_y}{2m}+V(x,y)-\mu\right]\tau_z + \Delta_B\, \sigma_x\nonumber\\
&+& \Delta_0\, \tau_x+\frac{\alpha}{\hbar}(p_x\sigma_y-p_y \sigma_x) \tau_z\; .
\end{eqnarray}
From left to right the contributions to Eq.\ \ref{eq.2.1} are: 
kinetic energy with 
$\vec{p}$ and $m$ the momentum and the effective mass, respectively; 
$V(x,y)$ is an electric potential representing the shape of the nanowire; 
$\mu$ the chemical potential; $\vec{\sigma}$ and $\vec{\tau}$ are vector
operators for spin and isospin (in electron-hole space) respectively;
$\Delta_B$, $\Delta_0$ and $\alpha$ represent the Zeeman, superconductivity 
and Rashba coupling energies, respectively. 
A sketch of the physical system is given in Fig.\ \ref{F2.1}.
We model rectangular shape potentials of lengths $L_x$ and $L_y$,
using Fermi functions of diffusivity $s_f$, such that the potential $V(x,y)$
vanishes inside the nanowire and takes a (large) value $V_0$ outside.

A natural unit system can be determined by the coupling constants $\alpha$, Planck's constant $\hbar$, and the effective mass $m$ of the semiconductor. The hamiltonian becomes dimensionless in the following unit system,
\begin{equation}
\label{eq.2.2}
E_{so}=\frac{\alpha^2 m}{\hbar^2},\quad L_{so}=\frac{\hbar^2}{\alpha m}.
\end{equation}
In these units the constants $\alpha$, $\hbar$, and $m$ are chosen equal to one. All the results
of the work  are given in these units, unless otherwise specified.

The superconductivity effect induces particle-hole symmetry, yielding symmetric energy eigenvalues with respect to the chemical potential. Moreover, superconductivity is responsible for an energy gap around the chemical potential. The Rashba spin-orbit coupling originates in the interaction between the electron spin and its own motion and,  finally, the Zeeman term allows us to drive the system into different regimes. Successive Majorana states for each transverse mode $n=1,2,\dots$ may appear when the Zeeman term overcomes critical values 
\begin{equation}
	\Delta_{B,n}^{(c)}=\sqrt{\mu_{n}^2+\Delta^2_0}\,
	\label{eq.2.3}
\end{equation}
where $\mu_n$ is an effective chemical potential for each transverse band.
Neglecting the Rashba mixing term ($\alpha p_y\sigma_x\tau_z$) each transverse 
mode leads to an independent Majorana that may coexist 
with other $n$ Majoranas
for sufficiently large Zeeman
fields.
In the limit of strong $\alpha$,
however, 
just a single zero mode effectively survives 
due to mode-mode interactions \cite{Tewari2}.
Besides, in a finite nanowire the scenario 
of sharp transition
boundaries is distorted due to the finite-size effect \cite{Lim,gun}.  

In this work we will focus on the first transition point, considering Zeeman energies
from zero up to a value not much exceeding $\Delta_{B,1}^{(c)}$.
The reader may notice that we have not considered magnetic fields in other directions different from $\hat{x}$.
The effects of tilting the magnetic field are discussed in Ref.\ \cite{osca2014} with the 
so-called projection rule for a unidimensional nanowire.
For magnetic fields in the $xy$ plane the Majorana mode delocalizes when the azimuthal angle
 exceeds a critical value. Moreover, if the magnetic field is out of the $xy$ plane, 
the formation of Majorana states is destroyed due to orbital effects \cite{Lim}.

\subsection{Diagonalization method}

We shall solve the Schr\"odinger equation with the above Hamiltonian,
\begin{equation}
\label{eq.2.4}
\mathcal{H}_{BdG}\Psi(x,y,\eta_\sigma,\eta_\tau)=E\Psi(x,y,\eta_\sigma,\eta_\tau)\; ,
\end{equation}
where the wave function variables are
the space coordinates $(x,y)$ $\in$ $\Re$, the spin $\eta_\sigma$ $\in$ $\lbrace\uparrow,\downarrow\rbrace$ and isospin $\eta_\tau$ $\in$ $\lbrace\Uparrow,\Downarrow\rbrace$.
We expand  in a
basis of eigenspinors for spin and isospin, $\chi_{s_\sigma}$ and $\chi_{s_\tau}$,
\begin{equation}
\label{eq.2.5}
\Psi(x,y,\eta_\sigma,\eta_\tau)=\sum_{s_\sigma, s_\tau} \psi_{s_\sigma, s_\tau}(x,y)\, \chi_{s_\sigma}(\eta_\sigma) \chi_{s_\tau}(\eta_\tau)\; ,
\end{equation}
with the quantum numbers $s_\sigma=\pm$ and $s_\tau=\pm$. It is fulfilled that
\begin{equation}
\label{eq.2.6}
\sigma_z \chi_{s_\sigma}(\eta_\sigma)=s_\sigma \chi_{s_\sigma}(\eta_\sigma)\; ,
\end{equation}
\begin{equation}
\tau_z\chi_{s_\tau}(\eta_\tau)=s_\tau\chi_{s_\tau}(\eta_\tau).
\end{equation}

Projecting equation (\ref{eq.2.4}) on $\langle s_\sigma s_\tau\vert$ we find 
the following system of equations for the components
$\psi_{s_\sigma, s_\tau}(x,y)$ of the wave function
\begin{eqnarray}
\label{eq.2.7}
&& \left[\left(\frac{p_x^2+p_y^2}{2m}+V(x,y)-\mu\right)s_\tau-E\right]\psi_{s_\sigma s_\tau}(x,y)
+\Delta_0\, \psi_{s_\sigma \bar{s}_\tau}(x,y)
\nonumber\\
&+&
\left[
\Delta_B(\cos\varphi-i s_\sigma \sin\varphi)
-\frac{\alpha}{\hbar	}s_\tau (i s_\sigma p_x+p_y)
\right]
\, \psi_{\bar{s}_\sigma s_\tau}(x,y)=0\; ,
\end{eqnarray}
where we use the notation $\bar{s}=-s$.
In order to solve this equation system with partial derivatives we use numerical techniques. We have discretized the space in a square lattice, where $N_x$ and $N_y$ are the number of points in each direction. The boundary conditions are simply vanishing of the wave function at the edges of the grid. We use finite differences to describe the partial derivatives and transform Eq.\ (\ref{eq.2.7}) into 
a sparse matrix that we diagonalize with standard routines \cite{Harwell}.

\begin{figure*}[t]
\begin{center}
	\includegraphics[scale=0.50]{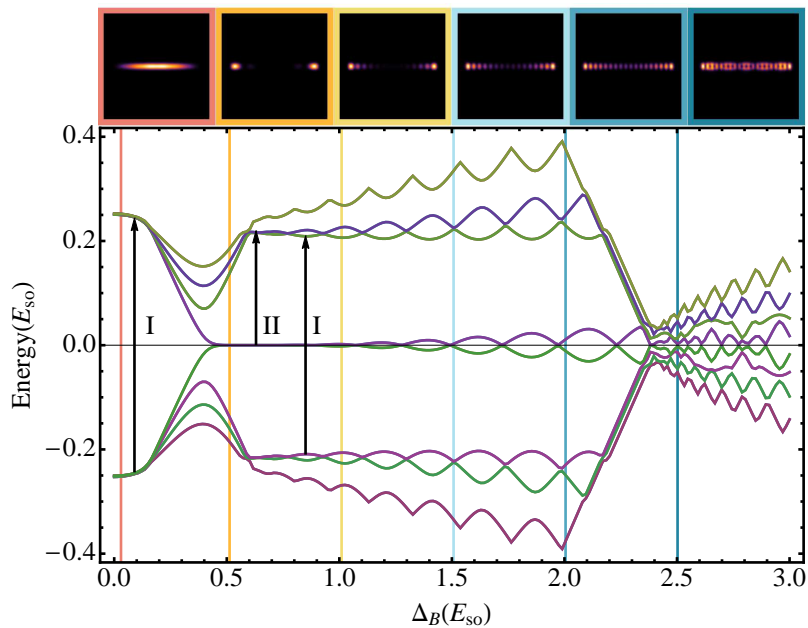} 
	\caption{Representation of the eight eigenvalues lying closer to zero energy as a function of Zeeman energy $\Delta_B$. We present also six different cuts (vertical lines) corresponding to different 
$\Delta_B$'s, showing the probability density of the lowest positive eigenvalue at the corresponding Zeeman energy. The color of each line and frame are matched to better indicate the value of $\Delta_B$ for each density.
Representative transitions of types I and II are indicated.
}
	\label{F2.2}
\end{center}
\end{figure*}

\subsection{Eigenvalues and densities}

In this section we are going to present the numerical solutions of the above Hamiltonian. These numerical solutions correspond to the following set of parameters. We assume a rather thin nanowire of 
$L_x=25 L_{so}$, $L_y=2 L_{so}$, maximum value of the potential outside the nanowire
$V_0= 5 E_{so}$, softness of the Fermi functions $s_f=0.1 L_{so}$.
The superconductivity energy is taken as $\Delta_0=0.25 E_{so}$ and, for simplicity, the chemical potential has been taken equal to zero, $\mu=0$.
We vary the Zeeman energy in order to monitor the emergence of zero modes. We emphasize that this set of parameters  will be used in the rest of the work as a representative case showing the emergence of 
Majorana physics.

Figure\ \ref{F2.2} shows the evolution of the eigenvalues as a function of the magnetic field.
We only display the  8 eigenvalues lying closer to zero energy, to avoid burdening the figure,
although we included up to 32 eigenvalues in the calculations discussed below. 
We also present in the same graph the probability density of the lowest positive eigenvalue for selected cases in order to see how the system evolves with increasing Zeeman energies.
For very low $\Delta_B$'s all the eigenvalues are nearly degenerate at two energies $E\approx\pm\Delta_0$ and the probability density is similar to that of a square well potential. 
When the Zeeman energy increases, the energy levels start to split, the gap becoming smaller and the first eigenfunction shows a quenched probability in the middle of the nanowire. 
For $\Delta_B\approx 0.42 E_{so}$ the gap closes completely and the Majorana forms, this fact can be appreciated because the wave function is characterized by two probability maxima well localized on the tips of the nanowire. If we continue increasing the Zeeman energy the gap reopens, but one state remains 'trapped' in the middle of the gap, the Majorana state.

The fact that the Majorana state lies in the middle of an energy gap effectively protects the Majorana mode from decoherence due to noise and disorder. Increasing further the magnetic field, the localized maxima spread on the nanowire more and more and 
an oscillation of the Majorana energy around zero is seen in Fig.\ \ref{F2.2} for $1.25 E_{so}<\Delta_B< 2.3 E_{so}$ due to the finite size effect.
For $\Delta_B\approx0.4E_{so}$ the
second transverse mode collapses and mode-mode interactions prevent  the existence of the zero mode
no longer.

\subsection{Particle-Hole symmetry}
\label{SEC2.2}

The main aim of this subsection is to provide an important result for the subsequent 
calculation of the cross section. 
For this reason, we introduce a symmetry operator $\Theta$ such that
\begin{equation}
\label{phS}
\Theta \mathcal{H}= -\mathcal{H} \Theta,
\end{equation}
where $\Theta$ is the time-reversal-plus-charge-conjugation operator (or time-charge inversion for short)
\begin{equation}
\Theta\equiv
-\sigma_y\tau_y\mathcal{K}=\left(\begin{array}{cccc}
0&0&0&1\\
0&0&-1&0\\
0&-1&0&0\\
1&0&0&0\\
\end{array}
\right)
\mathcal{K}
\end{equation}
with $\mathcal{K}$ is the conjugation operator.

Particle-hole symmetry, Eq.\ (\ref{phS}),
implies that if $\vert \Psi_E \rangle$ is an eigenstate with energy $E$ then $\Theta \vert \Psi_E \rangle$ is also an eigenstate but with energy $-E$, in agreement with the results of Fig.\ \ref{F2.2}.
Now we want to prove an important property with Bogoliubov-deGennes eigenstates: 
the matrix element of the momentum operator with particle-hole conjugate states is zero,
\begin{eqnarray}
 \label{2.10}
\langle \Psi_E \vert \vec{p}\, \vert  \Psi_{-E} \rangle &=& \langle \Psi_E  \vert \vec{p}\, \Theta \vert \Psi_E  \rangle=   \nonumber\\
&&
\!\!\!\!\!\!\!\!\!\!\!\!\!\!\!\!\!\!\!\!\!\!\!\!\!\!\!\!\!\!
 = -i\hbar \int^{+\infty}_{-\infty} (\psi^*_{\uparrow\Uparrow},\psi^*_{\uparrow\Downarrow},\psi^*_{\downarrow\Uparrow},\psi^*_{\downarrow\Downarrow}) \; \nabla \left(
\begin{array}{c}
 \;\;\;\psi^*_{\downarrow\Downarrow}\\
  -\psi^*_{\downarrow\Uparrow}\\
  -\psi^*_{\uparrow\Downarrow}\\
  \;\;\;\psi^*_{\uparrow\Uparrow}\\
\end{array}
\right) d^2r\nonumber \\
 &&
 \!\!\!\!\!\!\!\!\!\!\!\!\!\!\!\!\!\!\!\!\!\!\!\!\!\!\!\!\!\!
 = -i\hbar \int^{+\infty}_{-\infty} 
\nabla \left(
 \psi^*_{\uparrow\Uparrow} \psi^*_{\downarrow\Downarrow} -\psi^*_{\uparrow\Downarrow} \psi^*_{\downarrow\Uparrow} \right) d^2r = 0\; .
\end{eqnarray}
Due to boundary conditions the last integral vanishes, implying 
that transitions from $\vert \Psi_E \rangle$  to $\vert \Psi_{-E} \rangle$ 
will not contribute to the dipole optical spectrum.

\section{Dipole cross section formalism}
\label{SEC3.1}

We consider the action of a time-dependent perturbation of type
\begin{equation}
\label{eq.3.1}
\mathcal{H}=\mathcal{H}_{BdG}+\frac{e}{mc}\vec{A}(\vec{r},t)\cdot\vec{p}\; ,
\end{equation}
where $\vec{A}$ represents the vector potential of an electromagnetic wave given, in principle, by 
a wave packet
\begin{equation}
\label{eq.3.16}
\vec{A}(\vec{r},t)=\int^{+\infty}_{-\infty} A(\omega)  e^{-i\omega(t-\frac{\hat{n}\cdot \vec{r}}{c})}\cdot \:\hat{e} \: d\omega\; .
\end{equation}
In Eq.\ (\ref{eq.3.16}) the unit vector $\hat{e}$ indicates the polarization direction while
$\hat{n}$ corresponds to the direction of the wave propagation.
In the dipole approximation the latter becomes irrelevant and we may write the 
absorption cross section as 
\begin{equation}
\label{eq.3.28}
\sigma(\omega)=\frac{4 \pi^2 \alpha_{F}}{m^2} \sum_{k,s} 
\, \frac{\left| {\cal D}_{ks}  \right|^2}{\omega_{ks}}\,\delta(\omega-\omega_{ks})\, f_s\,(1-f_k)\; ,
\end{equation}
where $\alpha_{F}$ is the fine structure constant, $\hbar\omega_{ks}\equiv E_k- E_s$
and the dipole matrix element is
\begin{equation}
\label{dipme}
{\cal D}_{ks} = \langle k\vert \vec{p}\cdot \hat{e}\vert s\rangle\; ,
\end{equation}
and $f_{s,k}$ are the occupations of levels $s,k$Á as given by Fermi functions
with a given temperature $T$
\begin{equation}
f_s = \frac{1}{1+e^{E_s/k_B T}}\; .
\end{equation}

Equation (\ref{eq.3.28}) yields the cross section as a set of delta peaks at energies
$\hbar\omega_{ks}$. At strict zero temperature only transitions from negative to positive 
BdG eigenstates are allowed. Increasing $T$, other transitions smoothly activate due to the thermal 
modification of the Fermi occupations in Eq.\ (\ref{eq.3.28}).
The dipolar regime is justified noting that the wave length of the light is typically much larger than the nanowire and, therefore, the electromagnetic radiation affects homogeneously all the nanostructure.

The numerical computation of Eq.\ (\ref{eq.3.28}) requires 
considering all possible transitions and computing the corresponding cross section 
matrix elements. For this we need to know all the energy levels and their respective wave functions, obtained in the preceding section. 
The matrix elements are obtained from the grid-discretized wave functions, 
calculating the spatial integrals of the four components of the wave function
as grid sums.  
The action of the momentum operator is also obtained using finite difference derivatives. 
The polarization direction $\hat{e}$ determines which component of the momentum operator is contributing. We characterize the polarization with the azimuthal angle $\varphi$, varying from $\hat{x}$ for $\varphi=0^\circ$ to $\hat{y}$ for $\varphi=90^\circ$.

In order to give more physically intuitive plots of the cross section 
we have replaced the delta functions by
normalized Lorentzians 
\begin{equation}
\delta(\omega-\omega_{ks})
\to
\frac{\Gamma}{2 \pi}\frac{1}{(\omega-\omega_{ks})^2+\frac{\Gamma^2}{4}}\; ,
\end{equation}
where $\Gamma$ is the width of the Lorentzian. The final cross section is a 
smooth energy function, better showing the accumulation of absorption strength
at some energies.
We have assumed $\Gamma=0.05 E_{so}$, a value that is useful to distinguish the contributions of the different transitions while, at the same time, it is in reasonable agreement with the experimental capabilities of measurement. With an InAs semiconductor, for which $E_{so}=0.4\; {\rm meV}$,
the chosen $\Gamma$ 
would correspond to an experimental resolution of $20\; \mu{\rm eV}$, or what is the same $0.2 \; {\rm cm}^{-1}$. This is only slightly below the experimental resolution found in the bibliography 
on far-infrared absorption of semiconductor quantum dots \cite{merkt1993far,phillips1998far,demel1988far,gudmundsson1995bernstein,steinebach1997far,demel1991one}. 
There are experiments with better resolutions, but not focused on nanostructures \cite{RICHARDS:64}.

In practice we have to truncate the space of eigenvalues, numerically finding only those levels whose energy in absolute value is below a certain maximum energy. As a consequence, when considering
transitions from negative to positive energies we are including all the existing transitions only below
a certain cut off energy $\varepsilon_c$. 
In the results shown below we have included 32 eigenvalues (notice that Fig.\ \ref{F2.2}
only displays the 8 eigenvalues closer to zero energy) and the corresponding cut off 
is $\varepsilon_c\approx 0.65 E_{\rm so }$ for
all Zeeman energies $\Delta_B$. Therefore, 
to avoid artifacts due to missing transitions in our space of levels   
we will not consider absorption energies much higher than $\varepsilon_c$.

\section{Results}
\label{SEC3.2}

This section contains the main results of this work. We discuss the electromagnetic cross section of
2D Majorana nanowires focussing, specifically, on the signatures of the 
presence of zero modes. Thus, our main goal is to provide 
guides for the detection of zero modes with optical spectroscopy. 

\begin{figure}[t]
\begin{center}
         \includegraphics[scale=0.4]{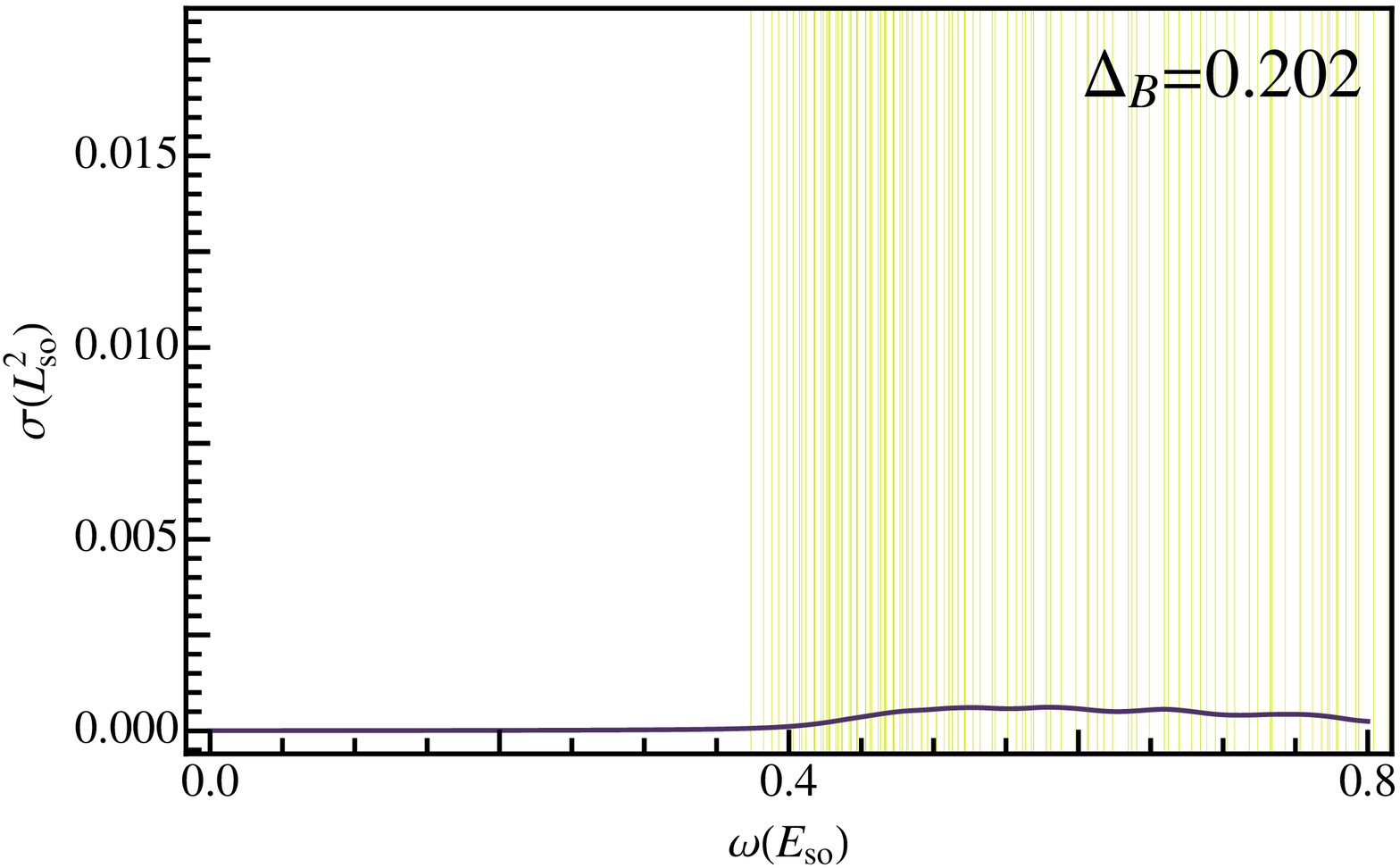} 
         \includegraphics[scale=0.4]{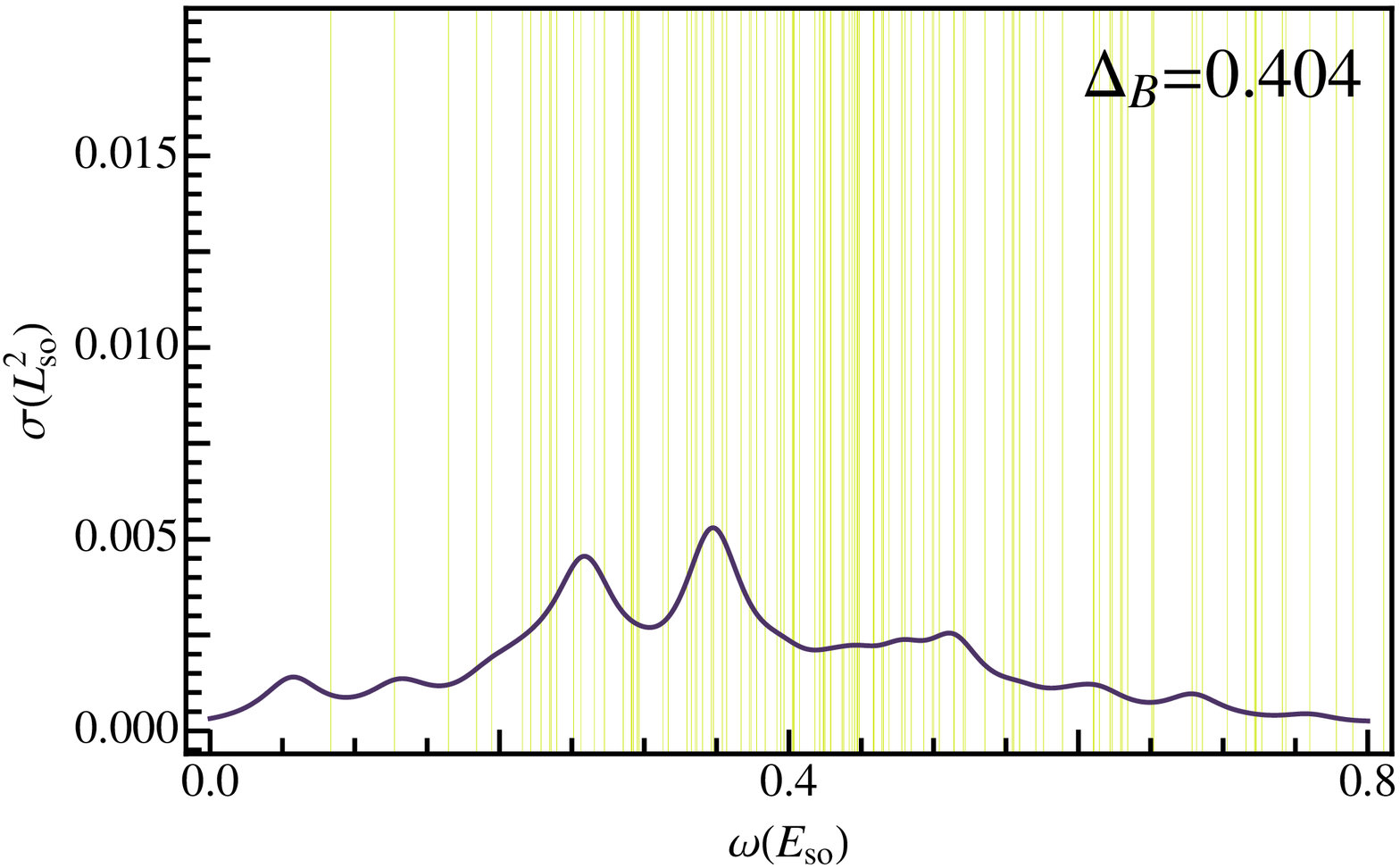} 
         \includegraphics[scale=0.4]{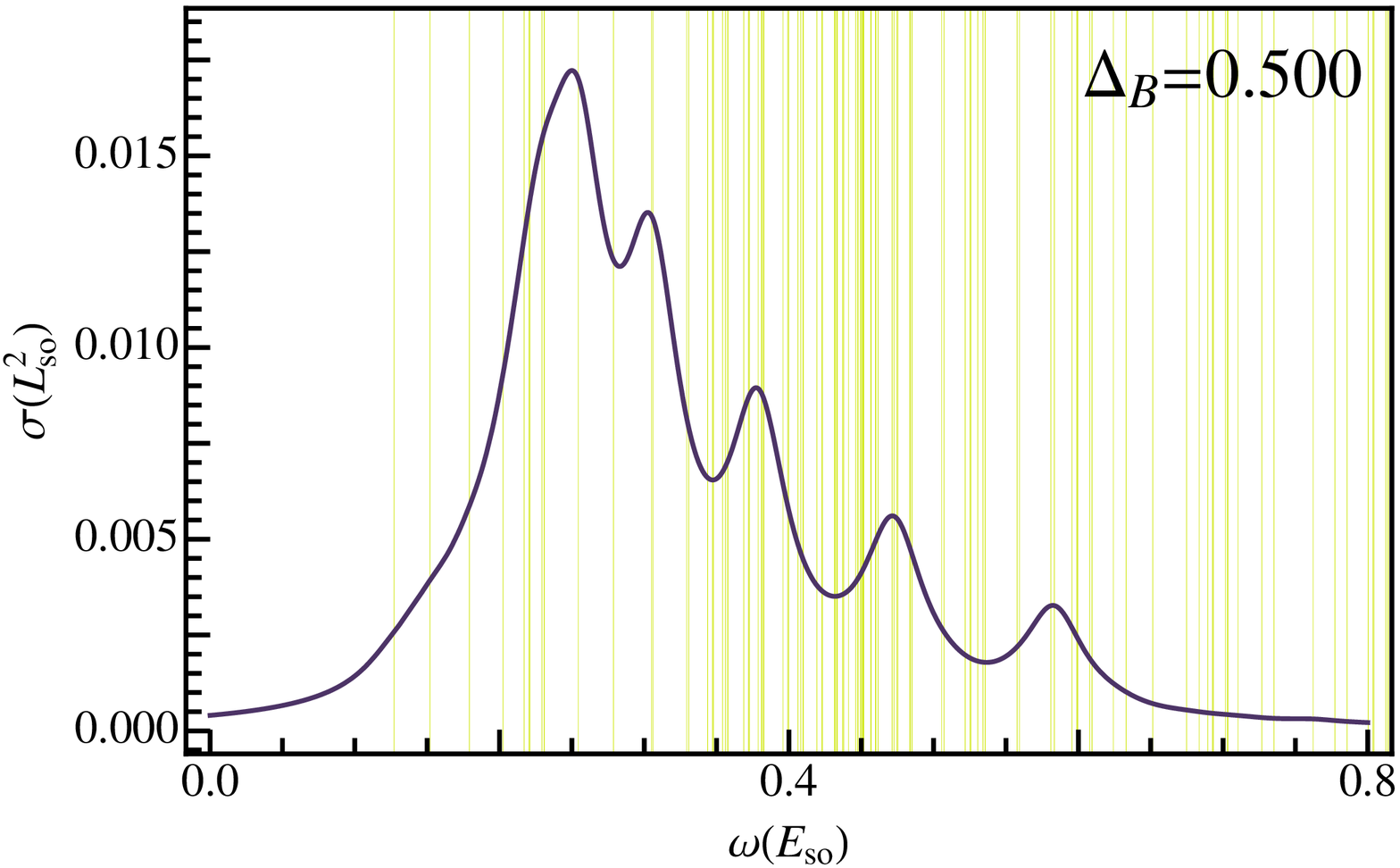} 
	\caption{Cross section for three selected values of the magnetic field. Thin vertical lines indicate the positions of all possible transitions from negative to positive energy levels, not counting forbidden transitions from $-E$ to $E$ conjugate levels. The calculations have been done with polarization along $\hat{x}$ and temperature $T=0.045\,{\rm K}$.}
	\label{FI3.2}
\end{center}
\end{figure}

\subsection{ Magnetic field dependence}
\label{SEC3.2.1}

The cross section is
characterized by transitions among
the system energy levels. For very small temperatures states at negative energies are occupied while
states at positive energies are empty.
For low values of $\Delta_B$ the energy spectrum is characterized by the 
presence of a wide 
gap centered on zero energy. This gap prevents any transition for energies lower than the corresponding energy jump ($\approx 0.5 E_{\rm so}$ in Fig.\ \ref{F2.2}). 
When the Zeeman energy increases, the gap diminishes and eventually closes 
for $\Delta_B\approx 0.42 E_{\rm so}$
when the lowest energy levels collapse into the Majorana state.

We may distinguish two types of transitions. The first group of transitions (type I) are those not involving the Majorana state and the second one (type II) are precisely those transitions from  the Majorana state to the rest.  Obviously, for $\Delta_B<\Delta_{B,1}^{(c)}$ only type I 
transitions  are present, while for 
$\Delta_B>\Delta_{B,1}^{(c)}$ both type I and type II are allowed.

The above scenario suggests the 
following criterion to infer the existence of a zero mode from experiments.
The Majorana mode causes the emergence of 
type II transitions, with the lowest type II transition lying at an energy which is exactly half that of the lowest type I transition. 
Notice that the lowest transition of type I coincides with the 
gap in the spectrum (not counting the Majorana state).
In the following we explicitly calculate the transition matrix elements to ascertain this scenario for different polarizations of the external field.

\subsubsection{Polarization along $\hat{x}$}

We analyze first the cross section for polarization 
along $\hat{x}$, the long axis of the nanowire,
for a low temperature ($T=0.045\; {\rm K}$). 
In Fig.\ \ref{FI3.2} we plot the computed cross sections for three selected Zeeman fields.
Each panel also shows 
with thin vertical lines
the positions of all  
possible transitions from negative to positive energy levels, not counting the 
forbidden transition between $E$ and $-E$ conjugate states. 
The upper panel is for a low value of the Zeeman energy, when there is a clear gap, and we can see that the
cross section although it is rather small
concentrates as expected in an energy region with many transitions. 
In the intermediate panel we can appreciate that the gap is very much reduced, as compared to the preceding panel, since transitions for lower energies are increasingly allowed. 
The reader should notice a peak, the first one in energy, without a corresponding 
vertical line. 
This is a temperature effect as can be deduced from the analysis of the data. Although the temperature is rather low, when the gap closes 
the first and the second negative energy levels become thermally activated and a transition between them can appear. 
For the lower panel, as seen in Fig. \ref{F2.2}, there is Majorana state.
The manifestation in the cross section is not clear, since
the lowest transition
at $\omega\approx0.12 E_{so}$ involving the Majorana state (type II)
only creates a very slight low-energy shoulder on the cross section.

\begin{figure}[t]
\begin{center}
	\includegraphics[scale=0.5]{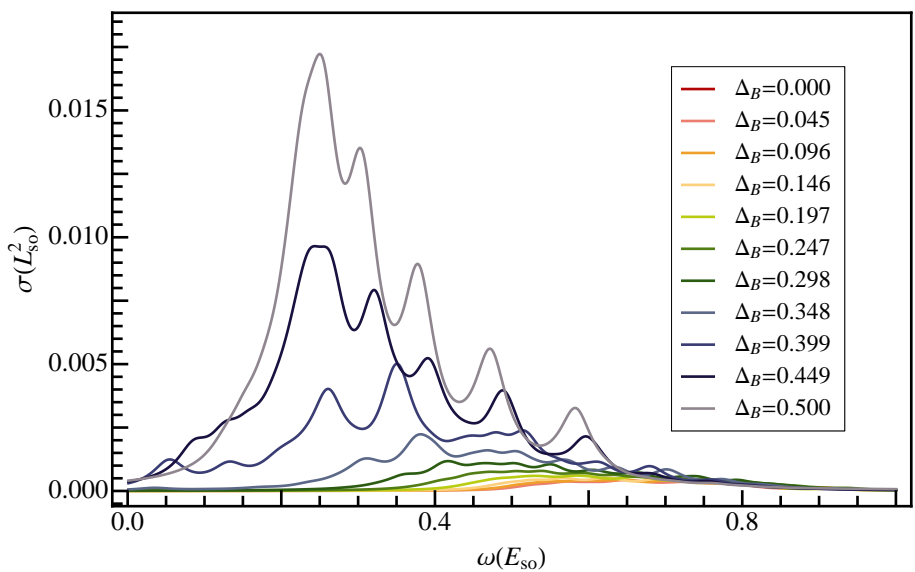} 
	\caption{Cross section with polarization along $\hat{x}$ and temperature $T=0.045\,{\rm K}$
	for selected values of the magnetic field (in $E_{so}$ units).}
	\label{FI3.3}
\end{center}
\end{figure}

In Fig.\ \ref{FI3.3} we have 
superposed in the same plot different cross sections to better see how this quantity evolves. The cross section for low Zeeman energies is very small, almost flat compared with the results of high Zeeman energies. 
We have checked that we include enough levels in the analysis and
 the effects of the cut off are small. For $\Delta_B=0.449 E_{so}$ we can see clearly that there are transitions for energies lower than the energy gap (the energy gap for type I transitions usually coincides with the highest  peak of the cross section), but these type II transitions leave a minor 
 fingerprint in the absorption spectrum as they are rapidly hidden by much higher peaks originating from transitions across the gap (type I).
As the magnetic field is increased there is a tendency to decrease the energy of the peaks, due to the gap closing.
There is also a clear tendency to increase the height of the absorption peaks as the gap closes. 
We conclude that for $\hat{x}$ polarization type II 
transitions are not easily visible.

\subsubsection{Polarization along $\hat{y}$}

\begin{figure}[t]
\begin{center}
	\includegraphics[scale=0.4]{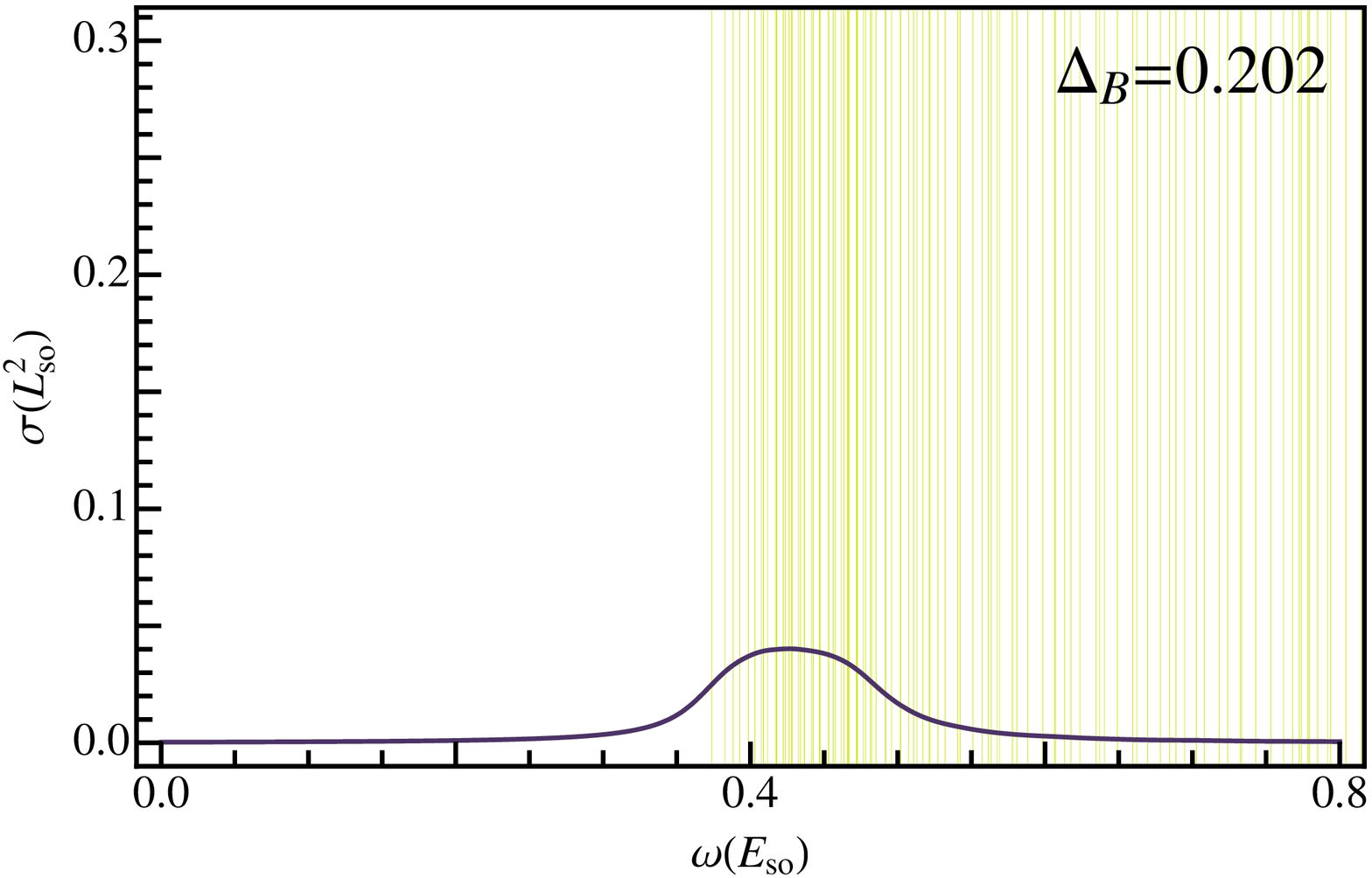} 
        \includegraphics[scale=0.4]{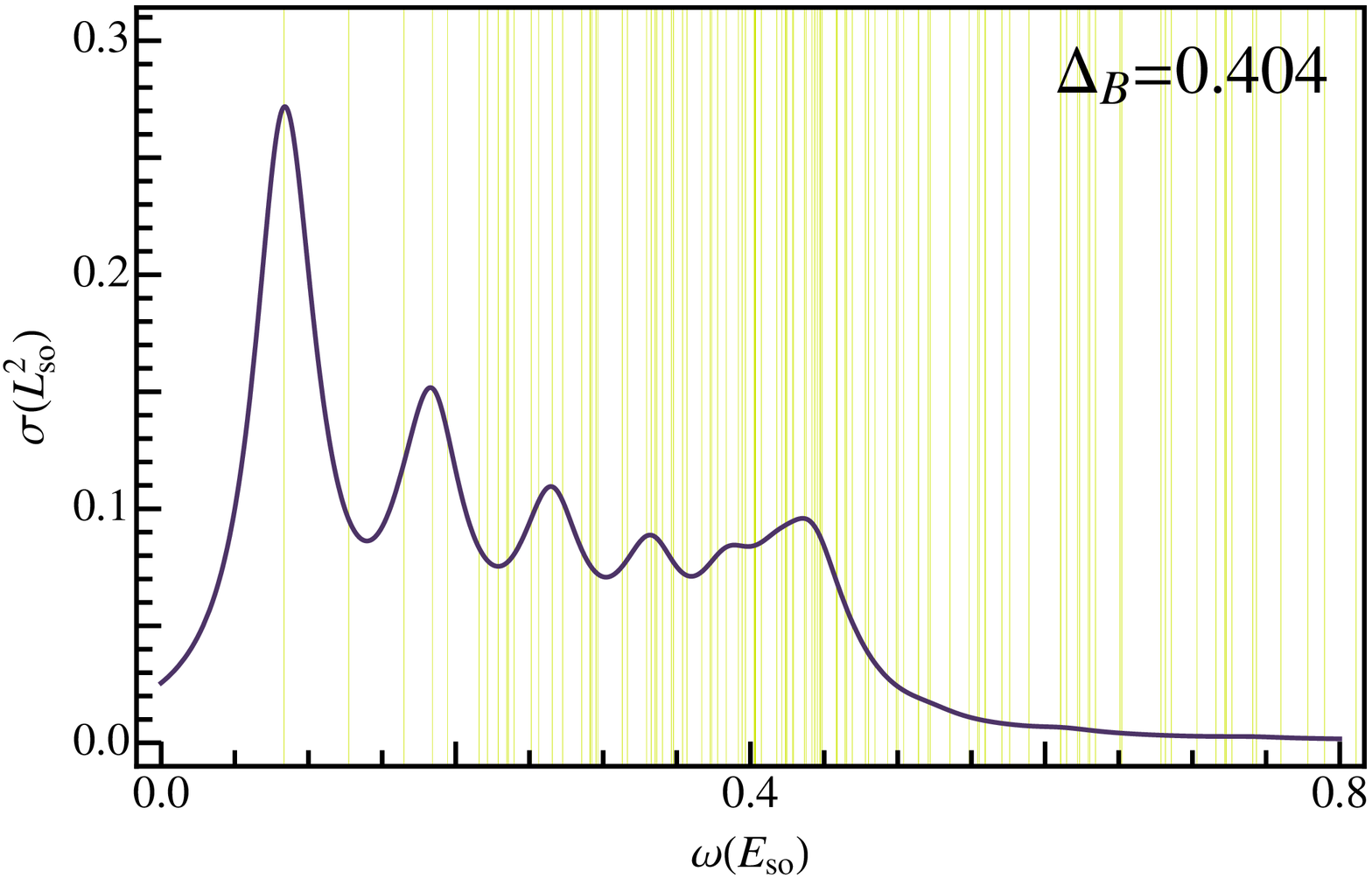} 
	\includegraphics[scale=0.4]{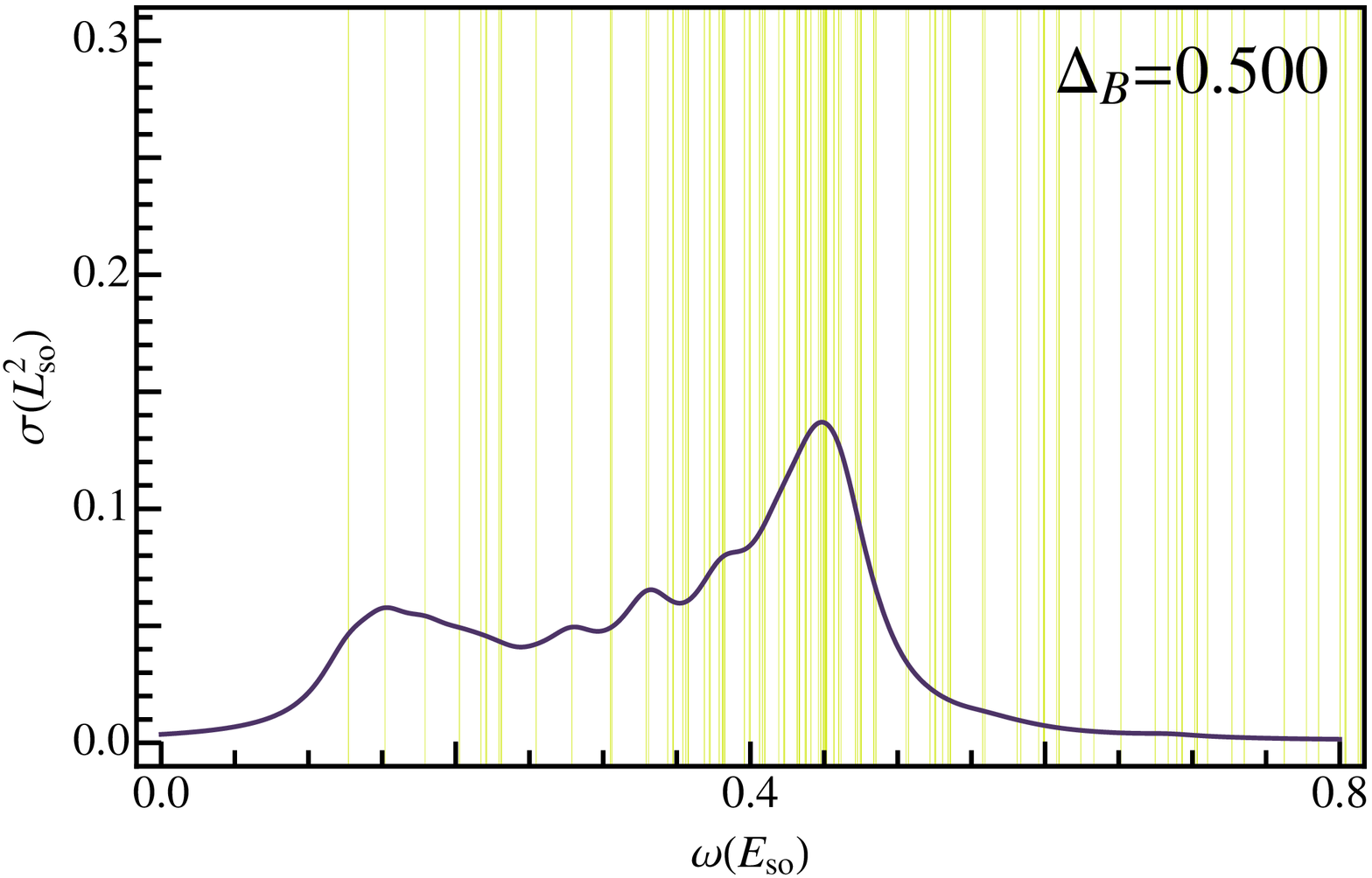} 
	\caption{Three different representations of the cross section for different values of magnetic field with the possible transitions explained in section \ref{SEC3.2.1}. The results are for $\hat{y}$ polarization and temperature $T=0.045\,{\rm K}$.}
	\label{FI3.5}
\end{center}
\end{figure}

Contrary to the preceding case of polarization along $\hat{x}$, the manifestations of the presence of a Majorana state can be clearly seen with $\hat{y}$ polarization. Figure \ref{FI3.5} shows these effects. In the upper panel the presence of the gap forbids low energy transitions. There is clear
tendency to decrease the cross section towards high energies, well below the cut off $\varepsilon_c\approx 0.65 E_{so}$.
In the intermediate panel the gap is almost closed and there are transitions at low energies. The lowest transition for this magnetic field achieves the maximum of the cross section.
The lower panel corresponds to the configuration having the Majorana state
and it can be clearly seen that type II transitions yield now a clear peak at low energy.
The first transition is around $\omega \approx 0.12 E_{so}$ while the gap for type I transitions is at $\omega \approx 0.25 E_{so}$. This lower transition must involve the Majorana state. 

It is remarkable  that with $\hat{y}$ polarization low energy type II transitions are strong enough to confirm the presence of a Majorana state. Furthermore, the cross section is such that the two groups of transitions can be clearly identified. 
We can also see in this third panel that around $\omega \approx 0.45 E_{so}$ there is a dense group of transitions, associated with the maximum absorption. This is again different from the case of $\hat{x}$ polarization, where the denser group is not the most absorbing.

\begin{figure}[t]
\begin{center}
	\includegraphics[scale=0.5]{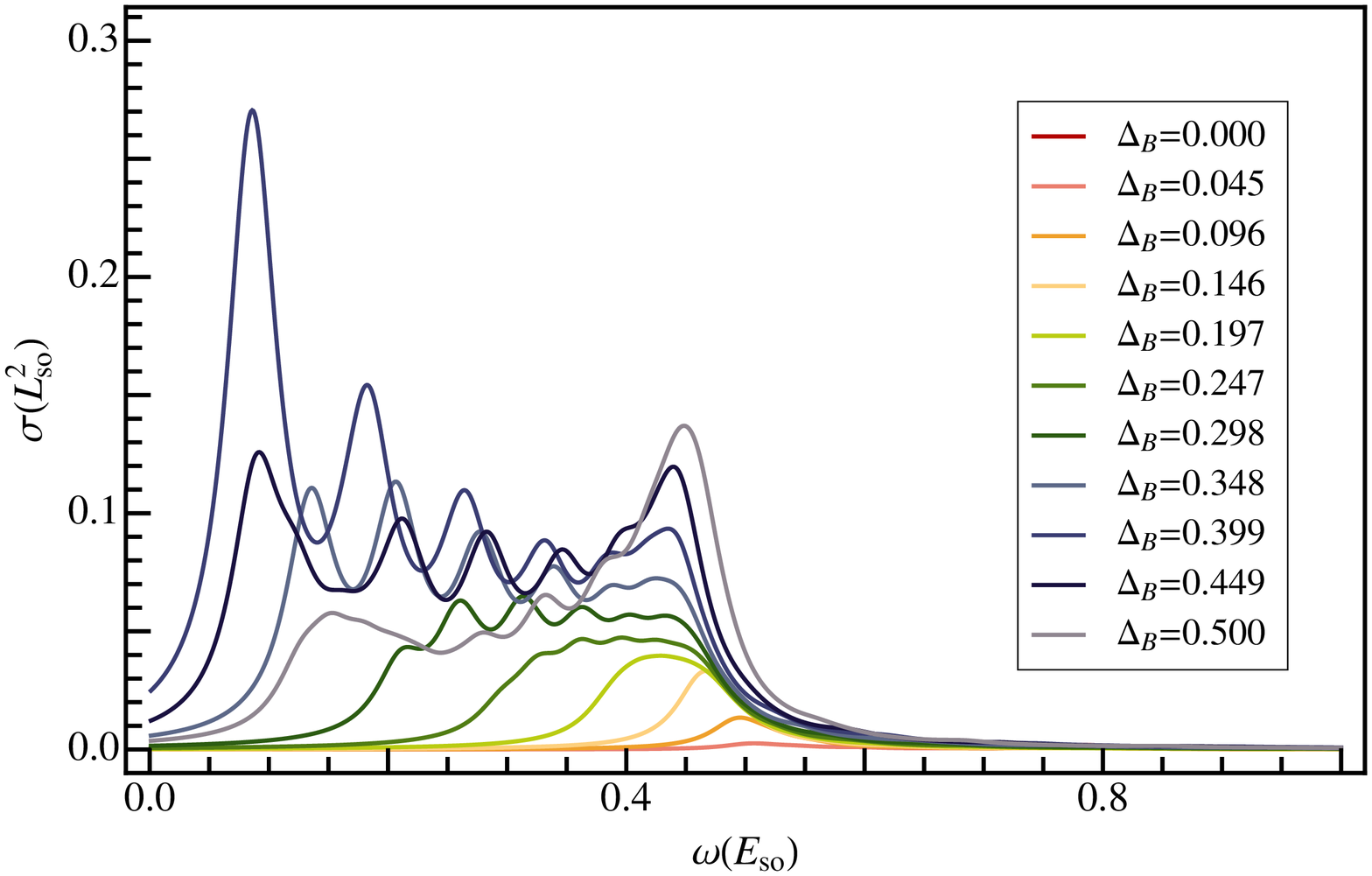} 
	\caption{Cross sections for selected magnetic fields  with polarization along $\hat{y}$ axis and temperature $T=0.045 K$.}
	\label{FI3.6}
\end{center}
\end{figure}

In Fig.\ \ref{FI3.6} we superpose the cross sections for varying  Zeeman energies in order to emphasize 
the variation with the closing of the gap. 
Moreover we can see from this figure that there is no peak for energies higher than $\hbar\omega \approx 0.5 E_{so}$, even though the set of eigenvalues contains transitions at such energies as shown by the  vertical lines in Fig.\ \ref{FI3.5}. We remind that all existing transitions for $\omega<\varepsilon_c\approx 0.65 E_{so}$ are included.
The interesting thing is that this high-energy behaviour remains constant with magnetic field, contrary to the $\hat{x}$ polarization case.
We understand this as a result of the 
subband grouping of the states in the finite system.

Physically, the differences between $\hat{x}$ and $\hat{y}$ polarizations can be attributed to the 
dipole selection rules with $p_x$ and $p_y$ operators.
A two dimensional system with strong confinement in the $y$ direction and a lower confinement in the $x$ direction has its levels approximately organized in groups $\varepsilon_{n_xn_y}$, such that 
changing $n_x$ we change energy level within a group, while changing $n_y$ we change to 
another group. Therefore when $n_y$ is changed the transition energy is higher than the corresponding change in $n_x$. The $\hat{x}$ polarization matrix elements  involve $p_x$ and this
operator changes the longitudinal mode of the wave function while, on the other hand, the  $\hat{y}$ polarization involves matrix elements of $p_y$ and requires transitions that change the transversal modes. In conclusion, the $\hat{x}$ polarization cross section involves transitions within the same group while 
the $\hat{y}$ polarization induces transitions to a different group. From this point of view it is not a surprise that the truncation of the set of eigenvalues affects differently the $\hat{x}$ and $\hat{y}$ polarized cross sections at high energies. This effect also explains that at low energies, the $\hat{y}$ polarization 
type II transitions are not dominated by other low energy type I transitions.

\subsection{Polarization rotation}

\begin{figure}[t]
\begin{center}
	\includegraphics[scale=0.4]{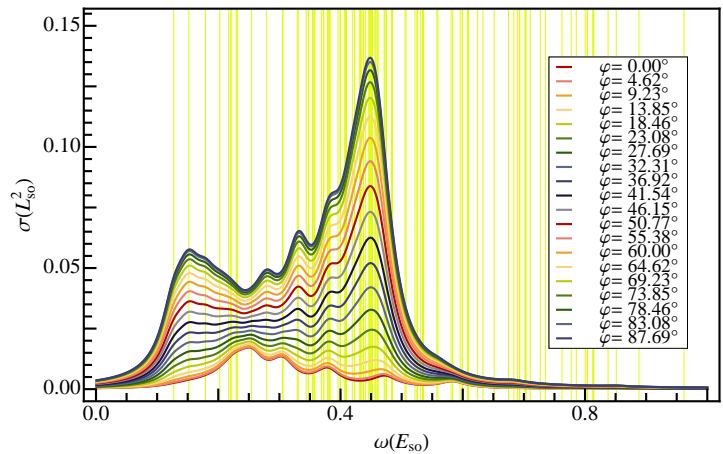} 
	\caption{Cross section representations for different values of the polarization angle $\varphi$ (in degrees). The polarization angle changes from $\hat{x}$ polarization to $\hat{y}$ polarization. These 
results are
for a constant magnetic field $\Delta_B=0.5 E_{so}$ and temperature $T=0.045 K$.}
	\label{FI3.9}
\end{center}
\end{figure}

We present now the explicit dependence on polarization direction at constant magnetic field when the Majorana is well formed. In Fig.\ \ref{FI3.9} we display the cross section for different values of the polarization angle $\varphi$, changing continuously from $\hat{x}$ to $\hat{y}$ direction. We have superimposed the possible transitions in order to see which polarization activates which transition.
The lowest curve, corresponding to $\hat{x}$ polarization is the same curve presented in the third panel of Fig.\ \ref{FI3.2}. The induced transitions by $p_y$ start to grow when the polarization angle increases and the $y$ component of the polarization vector becomes larger. Eventually, the induced transitions for $\hat{x}$ polarization fade away when the polarization angle reaches $\varphi=90^\circ$. We can see clearly that the polarization along the $\hat{y}$ axis enhances type II transitions
at low energy, a characteristic of the Majorana state.

One of the problems we should face in the experimental analysis of the absorption
cross section is that we cannot distinguish if the absorption spectrum 
has a contribution 
at lower energies
than the gap if we do not know the value of the gap. 
Figure \ref{FI3.9} suggests using the polarization effect.
Since for polarization along $\hat{x}$ absorption starts at the energy of the gap we can extract the corresponding value from this experimental information. After that, we can change to polarization along $\hat{y}$ and, if lower transitions appear, we can infer the presence of a Majorana state. 

\subsection{Temperature dependence}

In the preceding subsections we were considering a low temperature regime. For higher temperatures the thermal occupations start to play an important role in the transitions, with the possible transitions being not 
only restricted from negative energy to positive energy. We shall discuss the temperature dependence starting from low temperature $T=0.045\, {\rm K}$ and increasing until $T=4.523\, {\rm K}$ for the two extreme polarizations.
 
\subsubsection{Thermal ${\hat{x}}$ polarization}

\begin{figure}[t]
\begin{center}
	\includegraphics[scale=0.5]{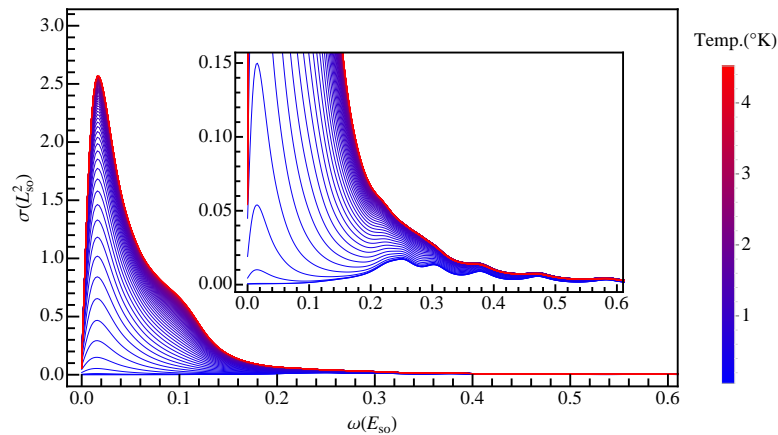} 
	\caption{Evolution of the cross section as a function of temperature for polarization along $\hat{x}$ and constant magnetic field $\Delta_B=0.5 E_{\rm so}$. Different curves of the cross section are presented, the temperature of each one being determined by the color according to the legend. The first panel shows all the curves while the second one zooms in the low temperature part.}
	\label{FI3.10}
\end{center}
\end{figure}

In Fig.\ \ref{FI3.10} we can see the temperature dependence for $\hat{x}$ polarization. In the upper panel we can see all the spectra while the lower one zooms in the lowest curves. For this polarization the temperature effects are very strong with all the signatures of the low temperature cross section being washed out already for $T\approx 4 {\rm K}$. It is not a surprise that a quantum effect is 
weak against temperature. In the present case the situation becomes dramatic due to the 
large weight of the low energy transitions induced by the $p_x$ operator.  

\begin{figure}[t]
\begin{center}
	\includegraphics[scale=0.5]{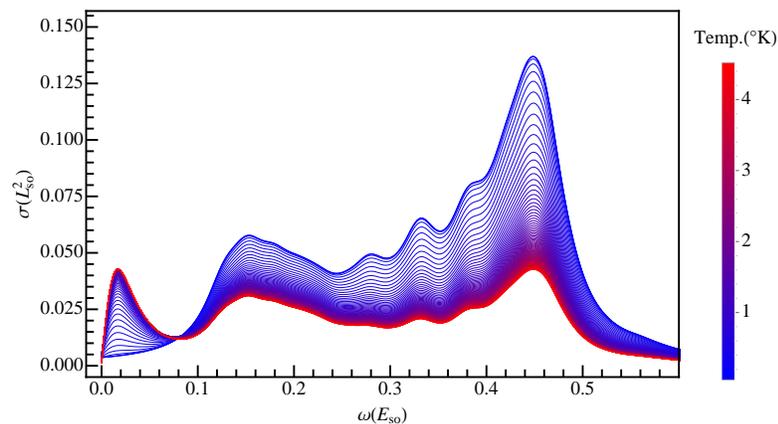} 
	\caption{Same as Fig.\ \ref{FI3.10} for $\hat{y}$ polarization.}
	\label{FI3.11}
\end{center}
\end{figure}

\subsubsection{Thermal $\hat{y}$ polarization}

A qualitatively different behavior is seen for the $\hat{y}$ polarization (Fig.\ \ref{FI3.11}). 
When the temperature increases in the same interval as before some low energy transitions appear, but these transitions do not cover the low temperature features of the cross section. The intensity of the
zero-temperature 
transitions diminishes with temperature but the
absorption feature at $\omega\approx 0.15 E_{so}$ due to
type II transitions remains clearly visible. This protection is
again due to the $\hat{y}$ polarization not allowing low energy transitions within the same group 
of states.
This is a remarkable  result suggesting the manifestation of a Majorana state at relatively high temperatures. 

\subsection{Optical masks}

We have considered the influence of optical masks hiding parts of the nanowire from the light.
Depositing an inert optical mask on the nanowire can be achieved, in principle, with lithographic techniques. 
We model the  mask effect simply adding a space dependent factor to the dipole matrix element of 
Eq.\ (\ref{dipme}), such that the integrand of the matrix element vanishes under the mask while it is unaffected in the 
rest of space. This way we want to check whether the absorption peaks, particularly the 
Majorana one,  strongly depend on the position of the mask or not.

\begin{figure}[t]
\begin{center}
	\includegraphics[scale=0.4]{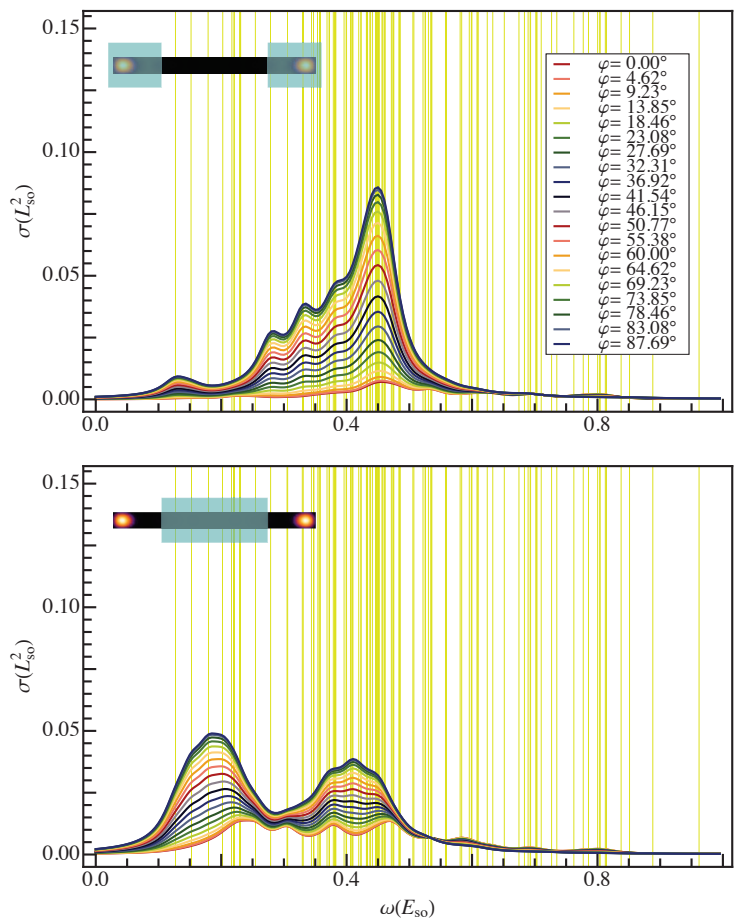} 
	\caption{Same as Fig.\ \ref{FI3.9} when a mask covers the nanowire ends (upper)
	and the nanowire center (lower). The position of the masks with respect to the 
	Majorana end states is indicated by the insets.}
	\label{FI3.12}
\end{center}
\end{figure}

Figure \ref{FI3.12} shows the evolution of the absorption with the polarization direction
when the mask covers the nanowire ends (upper) and the central part (lower). The results 
show that the two main absorption regions for $\hat{y}$ polarization are affected in opposite ways by the 
mask. Covering the nanowire ends quenches the absorption at low energies, while 
covering the center quenches the higher energy peaks. This behavior nicely confirms 
the expectations that Majorana modes are excited on the ends, due to their localization
properties. Our results suggest that space selective optical excitation 
is an effective way of probing the 
localization of Majoranas in a nanowire. 

\section{Conclusions}
\label{conclusions}

We calculated the absorption cross section of a 2D Majorana nanowire to a dipole field,
focussing on the absorption signatures of the zero mode. We suggested the emergence of 
low energy (type II) transitions when rotating from $\hat{x}$ to $\hat{y}$ polarization
as a clear fingerprint of the Majorana mode. These low energy feature is relatively robust 
against thermal activation of transitions that were forbidden at zero temperature. 
Finally, we suggested the use of optical masks to probe the localized character 
of the Majoranas on the nanowire ends.
As extensions of the present work, we think it is of  interest to consider
nanowires with cylindrical geometry as well as the influence of tilted fields on the 
optical absorption. 

\ack
This work was funded by MINECO-Spain (grant FIS2011-23526),
CAIB-Spain (Conselleria d'Educaci\'o, Cultura i Universitats) and 
FEDER. 

\section*{References}
\bibliographystyle{iopart-num}
\bibliography{TFM}

\end{document}